# Mirror contrastive loss based sliding window transformer for subject-independent motor imagery based EEG signal recognition

Jing Luo*[0000-0003-0465-2500], Qi Mao, Weiwei Shi, Zhenghao Shi, Xiaofan Wang, Xiaofeng Lu, Xinhong Hei*

Shaanxi Key Laboratory for Network Computing and Security Technology and Human–Machine Integration Intelligent Robot Shaanxi University Engineering Research Center, School of Computer Science and Engineering, Xi'an University of Technology, Xi'an, Shaanxi, China

luojing@xaut.edu.cn
heixinhong@xaut.edu.cn

**Abstract.** While deep learning models have been extensively utilized in motor imagery based EEG signal recognition, they often operate as black boxes. Motivated by neurological findings indicating that the mental imagery of left or right-hand movement induces event-related desynchronization (ERD) in the contralateral sensorimotor area of the brain, we propose a Mirror Contrastive Loss based Sliding Window Transformer (MCL-SWT) to enhance subject-independent motor imagery-based EEG signal recognition. Specifically, our proposed mirror contrastive loss enhances sensitivity to the spatial location of ERD by contrasting the original EEG signals with their mirror counterparts—mirror EEG signals generated by interchanging the channels of the left and right hemispheres of the EEG signals. Moreover, we introduce a temporal sliding window transformer that computes self-attention scores from high temporal resolution features, thereby improving model performance with manageable computational complexity. We evaluate the performance of MCL-SWT on subject-independent motor imagery EEG signal recognition tasks, and our experimental results demonstrate that MCL-SWT achieved accuracies of 66.48% and 75.62%, surpassing the state-of-the-art (SOTA) model by 2.82% and 2.17%, respectively. Furthermore, ablation experiments confirm the effectiveness of the proposed mirror contrastive loss. A code demo of MCL-SWT is available at https://github.com/roniusLuo/MCL_SWT.

## 1 Introduction

Brain-Computer Interface (BCI) establishes a direct communication channel between the brain and a computer, facilitating interaction and communication between cognitive processes and external devices, thus fostering the integration of biological and artificial intelligence [1-3]. Among various paradigms, Motor Imagery Brain-Computer Interface (MI-BCI) stands as a cornerstone, allowing users to manipulate external devices or perform specific tasks by mentally simulating movements. MI-BCI holds promising prospects in motor function rehabilitation [4]. Motor Imagery Electroencephalography (MI-EEG) captures EEG signals during motor imagery tasks, representing non-invasive, endogenous brain activity characterized by user-friendly operation, simplicity, flexibility, non-invasiveness, and minimal environmental requirements [5]. Accurate recognition of EEG signals is paramount for the development of robust subject-independent Motor Imagery (MI) Brain-Computer Interface (BCI) systems.

Presently, the primary focus of researchers in the realm of motor imagery EEG signal recognition algorithms lies in single-subject BCI systems, necessitating individual modeling of the target subject and yielding fruitful research outcomes [6]. However, these algorithms typically entail individualized calibration procedures, involving the collection of sufficient individual EEG signals and model adjustments [7]. This augments system complexity and calibration duration, thereby diminishing system usability and convenience. In contrast, subject-independent BCI systems endeavor to tackle the issue of inter-subject generalization, enabling BCI systems to better accommodate multiple subjects and expand the horizons of BCI technology applications [8]. Nonetheless, the spatial variance in event-related desynchronization/event-related synchronization (ERD/ERS) phenomena across different subjects presents significant research challenges [9]. Consequently, enhancing the model's ERD/ERS localization capability is imperative for bolstering subject-independent motor imagery EEG recognition performance.

Although deep learning models have been widely applied in recent years for MI based EEG signal recognition, they often function as black boxes and struggle to precisely localize ERD/ERS, a crucial factor in motor imagery recognition. Furthermore, while transformer-based approaches have demonstrated promising capabilities in extracting discriminative features from EEG signals [10], the computational complexity of the global multi-head self-attention mechanism in transformer models increases quadratically with the length of the input sequence. Currently, the majority of transformer models used in MI-EEG recognition employ short input sequences, thus limiting the temporal resolution of the extracted features

To address the issues mentioned above, we propose a mirror contrastive loss-based sliding window transformer (MCL-SWT) to enhance subject-independent motor imagery-based EEG signal recognition.

The main contributions of this paper are as follows:

(1) Motivated by neurological findings indicating that the mental imagery of left or right-hand movement induces event-related desynchronization (ERD) in the contralateral sensorimotor area of the brain, the MCL is proposed to enhance sensitivity to the spatial location of ERD by contrasting the original EEG signals with their mirror counterparts—mirror EEG signals generated by interchanging the channels of the left and right hemispheres of the EEG signals.

(2) A temporal SWT based subject-independent MI-EEG signal recognition model is proposed to achieve high time resolution of feature with affordable computational complexity. Specifically, the self-attention scores are calculated in temporal windows, and with the windows sliding along the temporal dimension of the EEG signal, the information from different temporal windows can interact with each other.

(3) The experimental results on subject-independent MI based EEG signal recognition demonstrate the effectiveness of MCL-SWT method in subject-independent MI-EEG classification tasks. Parameter sensitivity experiments have shown robustness of the MCL-SWT model, and the ablation study has validated the effectiveness of MCL.

## 2     Related work

### 2.1     CNN-based MI-EEG signal recognition method

In recent years, compared to traditional manually designed feature extraction methods, end-to-end feature extraction and classification approaches based on deep neural networks have demonstrated exceptional performance in the domain of motor imagery brain-computer interface (MI-BCI) [6]. Dai et al. introduced a hybrid-scale convolutional network architecture aimed at extracting temporal features of EEG signals across various convolutional scales for EEG-based motor imagery classification [11]. Yang et al. proposed a dual-branch 3D convolutional neural network for the three-dimensional representation of EEG data related to motor imagery, leveraging separate branches for temporal and spatial feature learning to circumvent mutual interference between these features. Furthermore, their framework introduced central loss to further enhance the decoding accuracy of motor imagery EEG [12]. Schirrmeister et al. delved into the impact of different CNN architecture designs on decoding MI-EEG signals, demonstrating superior performance of their proposed Deep ConvNet and Shallow ConvNet compared to alternative methods [13]. Lawhern et al. introduced a compact convolutional neural network dubbed EEGNet, showcasing its versatility across four BCI paradigms and its superior performance over other methods [14]. Mane et al. proposed a filter-bank convolutional network (FBCNet) for MI classification, leveraging multiple bandpass filters, deep convolutional layers for spatial information extraction, and innovative temporal aggregation techniques, outperforming existing methods on EEG signals from both healthy subjects and stroke patients [15]. Zhang et al. devised a weighted convolutional siamese network (WCSN) based on metric learning for feature representation of EEG signals, enhancing decoding accuracy by learning low-dimensional embeddings and implementing an adaptive training strategy to tackle non-stationarity between sessions [16]. Their method achieved significantly better decoding results on both limb neurorehabilitation and healthy subject datasets compared to state-of-the-art approaches. Hou et al. introduced a novel deep learning framework based on graph convolutional neural networks (GCNs) to enhance the recognition performance of raw EEG signals across various motor imagery tasks by capturing functional topological relationships of EEG channels [17]. Their approach involved constructing the Laplacian graph of EEG channels and employing GCNs-Net for feature extraction, followed by dimension reduction and final prediction through fully connected softmax layers.

### 2.2     Multi-subject MI-EEG signal recognition method

When implementing a single-subject brain-computer interface (BCI) system on a new subject, conducting experiments to collect EEG data for calibration becomes necessary.

This process is time-consuming and labor-intensive, significantly raising the practical application challenges of BCI systems. Consequently, researchers have begun exploring multi-subject motor imagery brain-computer interfaces.

Kwon et al. devised a pioneering framework based on deep convolutional neural networks for spectrum-space feature representation, tailored for multi-subject zero-calibration motor imagery brain-computer interface (MI-BCI). Leveraging a filter bank with multiple frequency bands in conjunction with mutual information and convolutional neural networks, this method constructs generalized features from diverse subjects and frequency bands, exhibiting remarkable performance on the Open BMI dataset [8]. Luo et al. introduced a twin-cascade softmax convolutional neural network for a multi-subject motor imagery BCI. To mitigate subject variability, they employed a cascaded softmax structure comprising subject identification and motor imagery recognition layers. Through joint optimization of subject identification and motor imagery recognition costs during model training, they achieved simultaneous subject identification and motor imagery recognition [18]. Hermosilla et al. developed a novel shallow convolutional neural network model for motor imagery classification, incorporating two convolutional layers for temporal and spatial feature extraction. Through single-subject and multi-subject experiments on three benchmark datasets, their approach surpassed state-of-the-art techniques [19]. Luo et al. proposed a shallow Transformer model for EEG signal decoding in motor imagery tasks. Utilizing multi-head self-attention layers with a global receptive field, they detected and utilized discriminative segments across multiple subjects EEG signals, enhancing classification accuracy. Furthermore, they improved performance through ensemble learning-based network structure and mirror EEG signal construction [5].

### 2.3     Attention mechanism-based MI-EEG signal recognition method

The attention mechanism allocates varying attention weights to different data or feature subsets, enabling the model to prioritize key areas, thereby acquiring detailed information about the target of interest while suppressing irrelevant information [20]. The self-attention mechanism calculates attention weights between each position and other positions to determine their significance in processing. It dynamically learns relationships across different positions in a sequence and addresses long-range dependencies [21]. Consequently, numerous researchers have integrated the attention mechanism with convolutional neural networks (CNNs) to construct models. Zhang et al. designed a convolutional recurrent attention model (CRAM), which uses CNN to encode the spatiotemporal information of EEG signal and establishes a recurrent attention mechanism to explore temporal dynamics between different time periods[22]. Altaheri et al. developed an attention-based temporal convolutional network (ATCNet) model for MI-EEG signal classification. ATCNet, a domain-specific and interpretable deep learning model, highlights valuable features in motor imagery EEG data using a multi-head self-attention mechanism and extracts advanced temporal features with a time convolutional network [23]. Wen et al. designed a CNN-based model architecture for end-to-end training and classification, incorporating a spatial-spectrum-temporal (SST) attention mechanism to adaptively extract the most expressive features from EEG data. Addi-tionally, they proposed a 3D Densely Connected Cross-Stage-Partial Network to segment extracted feature maps, reducing gradient loss and enhancing the model's representational capacity and robustness [24]. Amin et al. proposed a hybrid deep learning model architecture. Firstly, they employed the attention-inception convolutional neural network to extract spatial contextual features, which is crucial for learning the dynamic characteristics of EEG signal. Then, they utilized bidirectional long-short-term memory (Bi-LSTM) to learn temporal features[25]. Li et al. devised a temporal-spectral-based squeeze-and-excitation feature fusion network (TS-SEFFNet) for decoding motor imagery EEG signals. Their model incorporates a deep-temporal convolution block to extract high-dimensional information from EEG data, a multi-spectral convolution block for powerful spectral feature extraction, and a squeeze-and-excitation feature fusion block based on attention mechanism to enhance decoding performance [26]. Dong-Hee Ko proposed an attention-based deep learning approach to extract spatio-spectral features based on significant frequency bands for each subject. The method comprises three parts: extracting spatio-temporal features based on multiple frequency bands, utilizing sub-band attention to identify important frequency bands, and implementing an attention-based bidirectional long short-term memory network to extract time dynamic features [27]. Fan et al. proposed a new network structure for motor imagery EEG classification, called QNet. It includes a newly designed attention module (3D-Attention Module, 3D-AM) for learning attention weights of EEG channels, time points and feature maps. QNet uses a two-branch structure to learn more characteristics. After merging the dual branches, bilinear vectors are obtained. Finally, a fully connected layer is used as classifier[28]. Tao et al. proposed a novel solution called attention-based dual-scale fusion convolutional neural network (ADFCNN), which jointly extracts spectral and spatial information of EEG signal at different scales and provides new insights into integrating effective information from different scales through self-attention[29]. Li et al. introduced the depth-shallow attention multi-frame fusion network (DSA-MFNet) specifically designed for classifying motor imagery EEG signals. DSA-MFNet comprises the depth-shallow attention module for advanced feature extraction and the multi-frame fusion module for exploring inherent temporal variations in EEG data through multi-frame segmentation and recombination techniques [30].

In addition to embedding attention mechanism into convolutional neural network, transformer based models were proposed and applied in EEG decoding[31]. Besides the natural language processing field, successful vision model like Vision Transformer and Swin Transformer were proposed[32][33]. Google introduced a novel and streamlined network architecture known as the transformer model, specifically designed for sequence modeling and prominently featuring a self-attention mechanism. Departing from traditional convolutional and recurrent layers, the transformer model relies on a multi-head self-attention mechanism. This architecture has garnered significant success in natural language processing and machine translation tasks, demonstrating superiority in handling long-range dependencies and capturing global contextual information [31]. Subsequently, the computer vision community began adopting transformer models. Notably, the vision transformer model was proposed, which segments images into fixed-size patches and feeds them into an enhanced transformer model, outperforming CNN models. Leveraging multi-head self-attention mechanism and positional encoding, this model adeptly captures both global and local image information, enabling effective representation and processing of images with remarkable results [32]. Another innovative visual transformer model, the swin transformer, was introduced. This model strategically processes global information at lower resolutions using a local-partitioned attention mechanism and gradually integrates higher-resolution local information, thereby reducing computational and memory costs while maintaining accuracy. Additionally, it introduces a cross-window communication mechanism, focusing attention within offset windows to enhance computational efficiency and perception, facilitating information flow and feature representation within the network [33].

In summary, attention-related models have rapidly developed and demonstrated excellent representational capabilities, being widely applied in the field of brain computer interface, providing new insights for recognizing motor imagery EEG signal across multi-subject.

### 2.4    Loss function applied in MI-EEG signal recognition method

The loss function, serving as the optimization objective for training neural network models, has garnered significant attention and research focus within the realm of motor imagery recognition. Zhang et al. augmented the loss function by integrating regularization terms for acquired weights, employing squeeze-and-excitation modules to derive weights of EEG channels based on their contributions to EEG classification. They also devised an automated channel selection strategy. To fully exploit spatiotemporal characteristics, they proposed a convolutional neural network that notably outperformed traditional classification methods [34]. Autthasan et al. introduced a multi-task learning model termed MIN2Net, adept at extracting meaningful features from EEG data sans high-complexity preprocessing. Excelling in multi-subject motor imagery EEG classification, MIN2Net achieves end-to-end training through amalgamation of an autoencoder, deep metric learning, and supervised classifier. The autoencoder module aids in feature extraction from EEG data and furnishes discriminative patterns for diverse classes. The deep metric learning module aims to enhance feature discriminative power by refining distance measurement learning, while the supervised classifier module utilizes a standard softmax classifier to categorize latent vectors of input EEG signals. To derive a compact and distinctive latent representation from EEG signals, the model simultaneously minimizes reconstruction loss function, cross-entropy loss function, and triple loss function during optimization [35].

## 3     Method

This section provides a detailed description of the MCL and SWT model.

### 3.1    Notations and definitions

The original EEG signal is defined as $X \in R^{T \times C}$, where $T$ is the number of sampling points and $C$ is the number of EEG channels. The raw EEG signal $X$ serves as the input of the MCL-SWT model, with a batch size of $B$, resulting in an input data dimension of $B \times 1 \times T \times C$.

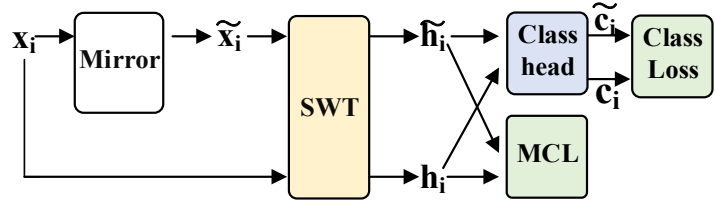

**Fig. 1.** SWT Model training framework based on MCL

## 3.2    Mirror contrastive loss

In EEG signals, imagining movements of the left or right hand can elicit the ERD phenomenon in the sensory-motor area on the contralateral hemisphere of the brain, and the ERS phenomenon on the same side of the brain. The accurate identification and localization of ERD/ERS are crucial criteria for MI classification. However, existing deep learning models only operate as black boxes and struggle to precisely locate the ERD/ERS phenomenon, leading to suboptimal results. To address this challenge and improve ERD/ERS localization ability of MI recognition model, we propose the MCL in this section. The framework of MCL based SWT model is illustrated in **Fig. 1**.

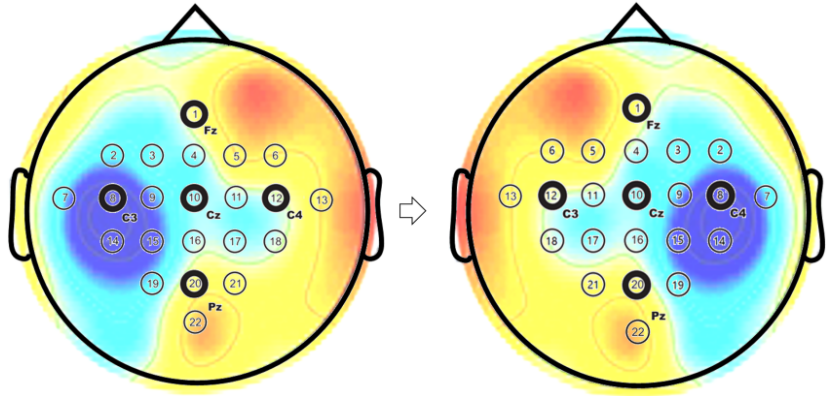

**Fig. 2.** Swapping left and right EEG channels to construct negative samples

**Mirror EEG signal.** To begin, the mirror EEG signal is generated by interchanging the channels of the left hemisphere and the right hemisphere of the EEG signals, as illustrated in **Fig. 2**. For instance, data from the original C3 channel is transposed to the C4 channel, and data from the C4 channel is transferred to the C3 channel. The electrode positions of the mirror EEG signals mirror those of the original EEG signals, hence the term 'mirror EEG signals'. As the ERD/ERS phenomenon in the mirror EEG signal manifests on the opposite side compared to the original EEG, and the corresponding left and right hand motor imagery label of the mirror EEG signal is designated as the opposite label of the original EEG signal. For instance, if the original EEG is labeled as left hand, then the label for the mirror EEG is right hand.

During the training phase, both mirror and original EEG signals are utilized in model training. By employing this sensible data augmentation technique, the number of training samples is doubled, thereby facilitating the model training process.

**MCL.** Consequently, the amalgamation of original and mirror EEG signals forms a negative sample pair, which can be leveraged in contrastive loss to bolster the capacity to pinpoint ERD/ERS phenomena through the contrast of deep features from negative sample pairs. Contrastive loss, as described in [40], optimizes the feature distributions during model training by encouraging the model to bring the features from positive sample pairs closer together while pushing features from negative sample pairs apart. The successful construction of positive and negative sample pairs is an essential factor in contrastive loss. While contrastive loss has been successfully applied in deep learning, it typically requires a large number of negative samples to effectively train the model and discern differences between samples.

In this paper, we propose a mirror contrastive loss to enhance the model's sensitivity to the location of ERD/ERS by contrasting the original EEG signals with their mirror EEG signal counterparts. This modification aims to improve the model's spatial awareness of ERD/ERS occurrences. Simultaneously, features less pertinent to the differences in EEG signals between the left and right sides of the brain are subdued. Furthermore, this approach effectively increases the number of negative sample pairs, thereby enhancing the effectiveness of contrastive learning loss.

In the MCL, the samples belonging to the same motor imagery task category are utilized as positive pairs, while the samples belonging to the different motor imagery task category are utilized as negative pairs. Specifically, Mirror Contrastive Learning (MCL) is applied to sample pairs from the original EEG signal and sample pairs from the original EEG signal and its mirror counterpart as:

$$L_d = w_o \sum_{i,j \in \text{OE}} g_{ij} D_{ij} + w_m \sum_{i \in \text{ME}, j \in \text{OE}} g_{ij} D_{ij} \qquad (1)$$

where $D_{ij}$ represents the Euclidean distance between sample pairs, consisting of sample $i$ and sample $j$. The variable $g_{ij}$ takes the value of 1 for positive sample pairs and -1 for negative sample pairs. Additionally, $w_o$ and $w_m$ denote the weights of each loss, while *OE* and *ME* refer to the original EEG signal set and mirror EEG set, respectively.

**Classification Loss.** This paper uses cross-entropy classification loss to train the temporal SWT EEG signal recognition model:

$$L_c = -\frac{1}{N} \sum_{i=1}^{N} y_i \ln\left(p_i\right) \qquad (2)$$

Therefore, the final loss function of the model training is the sum of the MCL and the classification loss:

$$L = L_c + L_d \qquad (3)$$

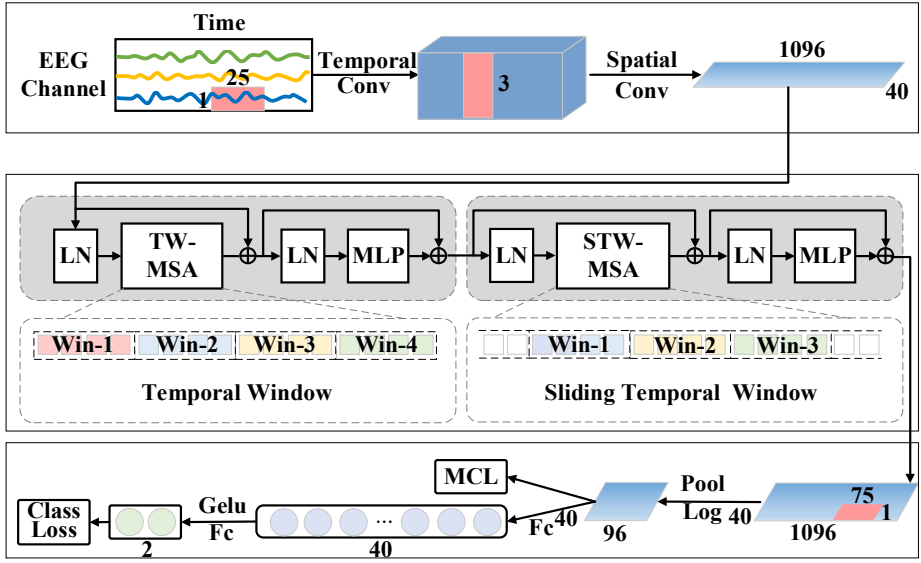

**Fig. 3.** Illustration of the overall architecture of the SWT model

### 3.3    SWT model

The overall architecture of the SWT model consists of three parts: the CNN-based feature extraction module, the sliding temporal window based multi-head self-attention module and the classification module, as shown in **Fig. 3**.

**CNN-based feature extraction block.** Initially, drawing inspiration from the bandpass filter utilized in the FBCSP algorithm [36], a temporal convolution $C_t$ with a kernel size of T×1 is used processing temporal information. Second, a spatial convolution $C_s$ with a kernel size of 1×C is applied to fuse information from each EEG channel. Finally, a batch normalization layer aiming to normalize the distribution of data in a batch is added [37]. The CNN-based feature extraction module can learn features with spatio-temporal information which can be described as

$$F = BN\big(C_s\big(C_t\big(X\big)\big)\big) \tag{4}$$

**Temporal Sliding Window based Multi-head Self-attention Module.** The temporal sliding window-based multi-head self-attention module is devised to establish a multi-head self-attention layer along the temporal dimension, aimed at capturing discriminative features of EEG signals with manageable computational load. This module comprises two stages, as depicted in Fig. 1. In the first stage, multiple self-attention layers are employed based on a temporal window, while the second stage facilitates information exchange between sequences in adjacent windows. Each stage is composed of two layer normalization (LN) layers, a temporal window multi-head self-attention layer, and a multi-layer perceptron (MLP) layer, integrated within a residual network structure. The sole discrepancy between these two stages lies in the division of temporal windows. A detailed description of each layer in this module is provided below:

(1) LN layer performs a layer normalization on the output feature $F$ of the CNN-based feature extraction module, normalizing the values along the feature dimensions of each sample[38].

(2) Temporal window multi-head self-attention layer (TW-MSA) is used to extract local temporal dependencies of input vectors in the time dimension, computing attention scores within a single window. It takes the layer-normalized feature as input, then segments it into a set of non-overlapping windows of size $M$. Self-attention scores computation is performed within each local window to enhance the model's perception of local information, and improve computational efficiency, as shown in Fig. 3. The window size in TW-MSA is empirically set to 8. The specific calculation steps of TW-MSA are as follows:

$$Q_i = W_i^Q \cdot LN(F) \tag{5}$$
$$K_i = W_i^K \cdot LN(F)$$
$$V_i = W_i^V \cdot LN(F)$$

$$h_i = Attention(Q_i, K_i, V_i) = \text{Softmax}\left(\frac{Q_i K_i^T}{\sqrt{d_k}}\right) V_i \tag{6}$$

$$O = concat(h_1, \cdots, h_H) W^o + F \tag{7}$$

where $h_i$ represents the i-th attention head, $H$ denotes the number of attention heads, $W_o \in R^{Hd_v \times d_{model}}$ represents the linear transformation matrix, $d_{model}$ is the dimension of the input embedding. $W_i^Q \in R^{d_{model} \times d_q}$, $W_i^K \in R^{d_{model} \times d_k}$, $W_i^V \in R^{d_{model} \times d_v}$ are the corresponding weight matrices. $d_q, d_k, d_v$ represent the dimensions of the query, key, and value respectively.

(3) MLP layer consists of a fully connected layer, a GELU[39] activation function layer, and another fully connected layer. Finally, the results of the input and output features are summed to produce the final output of the residual network structure.

$$A = O + FC\left(GELU\left(FC\left(LN(O)\right)\right)\right) \tag{8}$$

(4) Sliding temporal window multi-head self-attention layer (STW-MSA) calculates the attention scores in a cyclically shifted local windows for information interaction. As illustrated in **Fig. 4**, STW-MSA shifts temporal windows to the right with a step size of M/2, and calculates self-attention scores within the new temporal window.

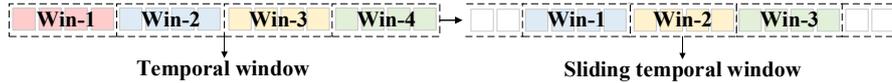

**Fig. 4.** Illustration of TW-MSA and STW-MSA

The sliding temporal window significantly reduces the computational complexity of the transformer model. While the computational complexity of the global multi-head self-attention mechanism is quadratic in relation to the length of the input sequence, as depicted in Equation (8), the computational complexity of TW-MSA is linearly related to the length of the input vector, as illustrated in Equation (9). Here, $L$ represents the length of the input vector, and $D$ denotes the input dimension. Consequently, TW-MSA exhibits a distinct advantage in terms of computational complexity.

$$\Omega(MSA) = 4LD^2 + 2L^2D \tag{9}$$

$$\Omega(TW-MSA) = 4LD^2 + 2M^2LD \tag{10}$$

**Classification module.** In the classification module, a square non-linear function is first applied. An average pooling layer is used to reduce the temporal dimension of features, followed by a logarithmic activation function. The EEG features are then inputted into a fully connected layer for classification. Finally, the softmax function is applied to compute the prediction probabilities. Assuming the output of the temporal multi-head self-attention module is $A$, the output of the classification module is:

$$Y = Softmax(FC(Log(AvgPool(Square(A))))) \tag{11}$$

Finally, the predicted probabilities for the mirror EEG signal corresponding to left MI and right MI are interchanged and then added to the predicted probabilities of the original EEG trial to obtain the final predicted probabilities [5]:

$$[Y_l, Y_r] = [Y_l^o + Y_r^o] + [Y_r^m + Y_l^m] \tag{12}$$

## 4    Experimental data and setup

### 4.1    Experimental data

The effectiveness of our proposed method was evaluated on BCI Competition IV dataset 2a and BCI Competition IV dataset 2b[42].

### 4.2    Experimental setup

In the experiment, the window size $M$ was set to 8. To divide the input feature vector of length $L$ into a whole number of non-overlapping windows, a 4.48-second EEG signal segment with 1120 sampling point from 0.5 s before MI cue onset to 3.98 s after MI cue onset was used. The bandpass filtering with a range of 4-38 Hz or 0-38 Hz was applied to the EEG signal, followed by channel-wise logarithmic sliding normalization. The preprocessed EEG signal was then input into the MCL-SWT model. During model training, Adam optimizer[43] was used as the optimization method with a weight decay parameter set to 0.05. As the new subject experiment  setup is applied, the two category and three channels consisting of C3, Cz, and C4 MI classification task as in dataset 2b is employed in this paper. Since the batch size is set to 100, the input data dimension would be $100 \times 1 \times 1120 \times 3$, and the output shapes and parameter quantities for each layer are shown in **Table 1**. The temporal convolution kernel T is set to 25, and the spatial convolution kernel C is set to 3. The weight coefficients $w_o$ and $w_m$ are empirically in (11) are set to 0.2 and 0.3, respectively.

**Table 1.** The output shape and parameters number of each layer

| Block | Layer | Output Shape | Param |
|---|---|---|---|
| Input | Input | [100，1，1120，3] | 0 |
| Feature Extraction Block | Temporal Conv | [100，40，1096，3] | 1040 |
| | Spatial Filter | [100，40，1096，1] | 4840 |
| | Feature Normalization | [100，40，1096，1] | 80 |
| | Rearrange | [100，1096，40] | 0 |
| (S)TW-MSA | Layer Norm | [100，1096，40] | 80 |
| | Query Projection | [100，1096，40] | 1640 |
| | Key Projection | [100，1096，40] | 1640 |
| | Value Projection | [100，1096，40] | 1640 |
| | Attention Score | [100，8，137，8，8] | 0 |
| | Projection | [100，1096，40] | 1640 |
| | Concatenate | [100，1096，40] | 0 |
| | Residual Add | [100，1096，40] | 0 |
| | Layer Norm | [100，1096，40] | 80 |
| | Linear | [100，1096，160] | 6560 |
| | GELU | [100，1096，160] | 0 |
| | Linear | [100，1096，40] | 6440 |
| | Residual Add | [100，1096，40] | 0 |
| | Layer Norm | [100，1096，40] | 80 |
| | Query Projection | [100，1096，40] | 1640 |
| | Key Projection | [100，1096，40] | 1640 |
| | Value Projection | [100，1096，40] | 1640 |
| | Attention Score | [100，8，137，8，8] | 0 |
| | Projection | [100，1096，40] | 1640 |
| | Concatenate | [100，1096，40] | 0 |
| | Residual Add | [100，1096，40] | 0 |
| | Layer Norm | [100，1096，40] | 80 |
| | Linear | [100，1096，160] | 6560 |
| | GELU | [100，1096，160] | 0 |
| | Linear | [100，1096，40] | 6440 |
| | Residual Add | [100，1096，40] | 0 |
| Classification Block | Average Pool | [100，40，69] | 0 |
| | Log | [100，40，69] | 0 |
| | Linear1 | [100，40] | 109640 |
| | Gelu | [100，40] | 0 |
| | Linear2 | [100，2] | 82 |

### 4.3    Dataset Division

This paper aims to investigate the classification performance of subject-independent motor imagery EEG signals, so the dataset is partitioned in a new subject setup as follows: when utilizing the training sessions of all 9 subjects in dataset 2a as the training set, the testing sessions (session 4 and session 5) of all 9 subjects in dataset 2b are employed as the testing set; conversely, when using the training sessions (session 3) of all 9 subjects in dataset 2b as the training set, the testing sessions of all 9 subjects in dataset 2a are utilized as the testing set. 5. Consequently, the trial data used for testing and training originate from entirely different subjects.

## 5    Results

### 5.1    Performance comparison of subject-independent MI recoginaton

This section evaluates motor imagery recognition performance of the MCL-SWT model on new subjects setup. Since the subjects in dataset 2a are entirely different from those in dataset 2b, the subjects in the testing set are new. As training typically converges within 500 epochs, the maximum number of epochs for training was set to 500 to ensure model convergence. Due to significant randomness affecting the test accuracy at a specific epoch, the evaluation metrics used in the experiment include: (1) Maximum test accuracy over 500 epochs (Max Accuracy); (2) Average test accuracy from epochs 401 to 500 (Average Accuracy); (3) Test accuracy at the epoch with the lowest training loss (Accuracy). **Table 2** presents the results in terms of "Accuracy/Kappa coefficient," with the maximum result highlighted in bold. The first row indicates the encoding format as "dataset - bandpass filter." For example, "2a-0 Hz" means the model was trained on dataset 2b and tested on dataset 2a using a 0-38 Hz bandpass filter for preprocessing. Five state-of-the-art models were compared in the experiment, including Shallow ConvNet, Deep ConvNet, EEGNet, FBCNet, and ATCNet. In additional, the ablation experiments are also conducted. The SWT and MCL-SWT indicate the SWT model without and with MCL, separately.

The experimental results in Table 2 lead to the following conclusions: (1) MCL-SWT achieved accuracies of 66.48% and 75.62%, surpassing the best state-of-the-art model by 2.82% and 2.17%, respectively. (2) The proposed temporal MCL-SWT model demonstrates superior performance across different datasets and bandpass filters. (3) MCL contributes to a further increase in the recognition accuracy and kappa value of SWT. (4) The proposed SWT model exhibits resistance to overfitting, as evidenced by the small gap between the maximum accuracy/Kappa and the average accuracy/Kappa of SWT. (5) The experimental results suggest that MCL-SWT possesses better inter-subject generalization ability.

**Table 2.** Classification performance (Accuracy and Kappa value) on the new subject

|  |  | 2a-0 Hz | 2a-4 Hz | 2b-0 Hz | 2b-4 Hz |
|---|---|---|---|---|---|
| Average Accuracy /Kappa | Shallow | 63.72/0.29 | 60.76/0.25 | 72.7/0.46 | 67.06/0.34 |
|  | Deep | 60.55/0.25 | 60.8/0.25 | 71.55/0.44 | 65.93/0.32 |
|  | EEGNet | 60.58/0.25 | 59.52/0.23 | 71.21/0.44 | 65.91/0.32 |
|  | FBCNet | 60.56/0.25 | 59.73/0.23 | 70.55/0.43 | 65.29/0.31 |
|  | ATCNet | 58.87/0.22 | 57.45/0.20 | 68.47/0.36 | 64.69/0.30 |
|  | SWT | 66.00/0.33 | 63.68/0.28 | 74.50/0.49 | 69.71/0.40 |
|  | MCL-SWT | **66.56/0.33** | **64.74/0.3** | **75.85/0.52** | **72.54/0.45** |
| Accuracy /Kappa | Shallow | 63.66/0.28 | 61.18/0.26 | 73.45/0.47 | 68.7/0.36 |
|  | Deep | 61.27/0.26 | 61.34/0.26 | 72.89/0.46 | 66.51/0.34 |
|  | EEGNet | 61.19/0.26 | 58.72/0.22 | 72.24/0.45 | 66.2/0.33 |
|  | FBCNet | 60.86/0.25 | 58.36/0.22 | 71.18/0.44 | 65.53/0.31 |
|  | ATCNet | 59.34/0.23 | 57.78/0.20 | 69.13/0.39 | 63.78/0.27 |
|  | SWT | 66.13/0.33 | 63.58/0.28 | 74.58/0.49 | 69.79/0.40 |
|  | MCL-SWT | **66.48/0.33** | **64.52/0.3** | **75.62/0.51** | **73.27/0.47** |
| Max Accuracy /Kappa | Shallow | 67.28/0.34 | 65.36/0.32 | 75.18/0.51 | 71.69/0.45 |
|  | Deep | 66.05/0.33 | 65.35/0.32 | 73.73/0.48 | 71.68/0.45 |
|  | EEGNet | 65.51/0.32 | 64.51/0.3 | 73.31/0.47 | 71.16/0.44 |
|  | FBCNet | 66.46/0.33 | 63.89/0.29 | 72.93/0.46 | 70.98/0.43 |
|  | ATCNet | 64.11/0.29 | 62.18/0.28 | 70.84/0.43 | 68.22/0.36 |
|  | SWT | 66.82/0.34 | 64.81/0.3 | 75.49/0.51 | 74.12/0.48 |
|  | MCL-SWT | **67.37/0.35** | **65.49/0.31** | **76.37/0.53** | **75.49/0.51** |

## 5.2    Sensitivity Analysis of Parameters

The number of temporal multi-head self-attention blocks and the number of heads in each self-attention block are two primary hyperparameters of the SWT model. This section investigates the impact of these hyperparameters on the model's performance, with the experimental results presented in **Table 3**. Increasing the number of heads or blocks leads to larger attention scores, resulting in a more complex model.

**Table 3.** The performance evaluation (2b-0 Hz) on different hyperparameter: (1) the head number of self-attention; (2) the number of temporal multi-head self-attention block

|  | 4 heads | 8 heads | 10 heads |
|---|---|---|---|
| 1 block | 74.71/0.49 | 74.5/0.49 | 74.73/0.49 |
| 2 block | 74.3/0.48 | 74.2/0.48 | 74.45/0.49 |
| 3 block | 73.16/0.46 | 73.67/0.47 | 73.26/0.46 |

The following conclusions can be drawn from the experimental results in Table 3: (1) In this experimental setup, the performance of the model deteriorates as the number of temporal multi-head self-attention blocks increases. This may be due to the limited number of training samples in motor imagery EEG, which is insufficient to fully train a model with multiple blocks; (2) The number of self-attention heads has a minimal impact on the model's performance; (3) The model exhibits robustness to various hyperparameters.

### 5.3    Model complexity analysis

The number of parameters and the inference time of the model are important indicators for assessing complexity. Therefore, this section compares these two aspects. Inference time is measured as the average time taken for 1000 runs. The experiments were conducted on Intel i7 10700K and NVIDIA GeForce RTX 3090. The specific experimental results are shown in **Table 4**.

**Table 4.** Model complexity analysis

|                    | Shallow | Deep | EEGNet | FBCNet | ATCNet | MCL-SWT |
|--------------------|---------|------|--------|--------|--------|---------|
| Parameter number/M | 10      | 268  | 3      | 3      | 37     | 155     |
| Inference time/ms  | 0.56    | 1.42 | 2.48   | 37.64  | 15.37  | 8.36    |

## 6    Conclusion

In this paper, we propose a Mirror Contrastive Loss (MCL) based on Sliding Window Transformation (SWT) model for subject-independent EEG signal recognition. By leveraging mirror EEG signals and MCL, the model aims to enhance sensitivity to the spatial location of ERD/ERS by contrasting the original EEG signals with their mirror counterparts. This modification is intended to improve the model's spatial awareness of ERD/ERS occurrences. Additionally, we introduce a temporal SWT that calculates self-attention scores in sliding windows, enhancing model performance with manageable computational complexity. The performance of MCL-SWT was evaluated on subject-independent motor imagery EEG signal recognition tasks. Experimental comparisons with state-of-the-art methods and ablation experiments demonstrate the superior performance of MCL-SWT. The proposed MCL serves as a general loss for MI-EEG recognition and can be integrated into various backbone networks. In future work, we aim to explore combining the proposed MCL with transfer learning in MI-EEG recognition.

**Acknowledgments.**
This work is jointly supported by the National Natural Science Foundation of China (Grant Nos. 61906152, 61976177, 62076201, 62376213 and U21A20524) and the Scientific          Research Program Founded by Shaanxi Provincial Education Department of China under Grant 23JK0556.

## References

1.  P. J. Benson, "Decoding brain-computer interfaces," Science, vol. 360, no. 6389, pp. 615-616, 2018.
2.  M. M. Shanechi, "Brain–machine interfaces from motor to mood," Nature neuroscience, vol. 22, no. 10, pp. 1554-1564, 2019.

3.  X. Tang, C. Yang, X. Sun, M. Zou, and H. Wang, "Motor imagery EEG decoding based on multi-scale hybrid networks and feature enhancement," IEEE Transactions on Neural Systems and Rehabilitation Engineering, vol. 31, pp. 1208-1218, 2023.

4.  S. Thomas, "Building a brain–computer interface to restore communication for people with paralysis," Nature Electronics, vol. 6, no. 12, pp. 924-925, 2023.

5.  J. Luo, Y. Wang, S. Xia, N. Lu, X. Ren, Z. Shi, and X. Hei, "A shallow mirror transformer for subject-independent motor imagery BCI," Computers in Biology and Medicine, vol. 164, pp. 107254, 2023.

6.  H. Altaheri, G. Muhammad, M. Alsulaiman, S. U. Amin, G. A. Altuwaijri, W. Abdul, M. A. Bencherif, and M. Faisal, "Deep learning techniques for classification of electroencephalogram (EEG) motor imagery (MI) signals: A review," Neural Computing and Applications, vol. 35, no. 20, pp. 14681-14722, 2023.

7.  F. Wei, X. Xu, T. Jia, D. Zhang, and X. Wu, "A multi-source transfer joint matching method for inter-subject motor imagery decoding," IEEE Transactions on Neural Systems and Rehabilitation Engineering, vol. 31, pp. 1258-1267, 2023.

8.  O.-Y. Kwon, M.-H. Lee, C. Guan, and S.-W. Lee, "Subject-independent brain–computer interfaces based on deep convolutional neural networks," IEEE transactions on neural networks and learning systems, vol. 31, no. 10, pp. 3839-3852, 2019.

9.  K. Zhang, N. Robinson, S.-W. Lee, and C. Guan, "Adaptive transfer learning for EEG motor imagery classification with deep Convolutional Neural Network," Neural Networks, vol. 136, pp. 1-10, 2021.

10. Z. Jia, Y. Lin, J. Wang, K. Yang, T. Liu, and X. Zhang, "MMCNN: A multi-branch multi-scale convolutional neural network for motor imagery classification." In Machine Learning and Knowledge Discovery in Databases: European Conference, ECML PKDD 2020, Ghent, Belgium, September 14–18, 2020, Proceedings, Part III (pp. 736-751). Springer International Publishing.

11. G. Dai, J. Zhou, J. Huang, and N. Wang, "HS-CNN: a CNN with hybrid convolution scale for EEG motor imagery classification," Journal of Neural Engineering, vol. 17, no. 1, pp. 016025, 2020.

12. L. Yang, Y. Song, X. Jia, K. Ma, and L. Xie, "Two-branch 3D convolutional neural network for motor imagery EEG decoding," Journal of Neural Engineering, vol. 18, no. 4, pp. 0460c7, 2021.

13. R. T. Schirrmeister, J. T. Springenberg, L. D. J. Fiederer, M. Glasstetter, K. Eggensperger, M. Tangermann, F. Hutter, W. Burgard, and T. Ball, "Deep learning with convolutional neural networks for EEG decoding and visualization," Human Brain Mapping, vol. 38, no. 11, pp. 5391-5420, 2017.

14. V. J. Lawhern, A. J. Solon, N. R. Waytowich, S. M. Gordon, C. P. Hung, and B. J. Lance, "EEGNet: a compact convolutional neural network for EEG-based brain–computer interfaces," Journal of Neural Engineering, vol. 15, no. 5, pp. 056013, 2018.

15. R. Mane, E. Chew, K. Chua, K. K. Ang, N. Robinson, A. P. Vinod, S.-W. Lee, and C. Guan, "FBCNet: A Multi-view Convolutional Neural Network for Brain-Computer Interface," arXiv preprint arXiv:2104.01233, 2021.

16. S. Zhang, K. K. Ang, D. Zheng, Q. Hui, X. Chen, Y. Li, N. Tang, E. Chew, R. Y. Lim, and C. Guan, "Learning EEG Representations With Weighted Convolutional Siamese Network: A Large Multi-Session Post-Stroke Rehabilitation Study," IEEE Transactions on Neural Systems and Rehabilitation Engineering, vol. 30, pp. 2824-2833, 2022.

17. Y. Hou, S. Jia, X. Lun, Z. Hao, Y. Shi, Y. Li, R. Zeng, and J. Lv, "GCNs-net: a graph convolutional neural network approach for decoding time-resolved eeg motor imagery signals," IEEE Transactions on Neural Networks and Learning Systems, 2022.

18. J. Luo, W. Shi, N. Lu, J. Wang, H. Chen, Y. Wang, X. Lu, X. Wang, and X. Hei, "Improving the performance of multisubject motor imagery-based BCIs using twin cascaded softmax CNNs," Journal of Neural Engineering, vol. 18, no. 3, pp. 036024, 2021.

19. D. M. Hermosilla, R. T. Codorniú, R. L. Baracaldo, R. S. Zamora, D. D. Rodriguez, Y. L. Albuerne, and J. R. N. Álvarez, "Shallow convolutional network excel for classifying motor imagery EEG in BCI applications," IEEE Access, vol. 9, pp. 98275-98286, 2021.

20. H. Larochelle, and G. E. Hinton, "Learning to combine foveal glimpses with a third-order Boltzmann machine," Advances in neural information processing systems, vol. 23, 2010.

21. A. Vaswani, N. Shazeer, N. Parmar, J. Uszkoreit, L. Jones, A. N. Gomez, Ł. Kaiser, and I. Polosukhin, "Attention is all you need," Advances in neural information processing systems, vol. 30, 2017.

22. D. Zhang, L. Yao, K. Chen, and J. Monaghan, "A convolutional recurrent attention model for subject-independent EEG signal analysis," IEEE Signal Processing Letters, vol. 26, no. 5, pp. 715-719, 2019.

23. H. Altaheri, G. Muhammad, and M. Alsulaiman, "Physics-informed attention temporal convolutional network for EEG-based motor imagery classification," IEEE Transactions on Industrial Informatics, vol. 19, no. 2, pp. 2249-2258, 2022.

24. Y. Wen, W. He, and Y. Zhang, "A new attention-based 3D densely connected cross-stage-partial network for motor imagery classification in BCI," Journal of Neural Engineering, vol. 19, no. 5, pp. 056026, 2022.

25. S. U. Amin, H. Altaheri, G. Muhammad, W. Abdul, and M. Alsulaiman, "Attention-inception and long-short-term memory-based electroencephalography classification for motor imagery tasks in rehabilitation," IEEE Transactions on Industrial Informatics, vol. 18, no. 8, pp. 5412-5421, 2021.

26. Y. Li, L. Guo, Y. Liu, J. Liu, and F. Meng, "A temporal-spectral-based squeeze-and-excitation feature fusion network for motor imagery EEG decoding," IEEE Transactions on Neural Systems and Rehabilitation Engineering, vol. 29, pp. 1534-1545, 2021.

27. D.-H. Ko, D.-H. Shin, and T.-E. Kam, "Attention-based spatio-temporal-spectral feature learning for subject-specific EEG classification." In 2021 9th International Winter Conference on Brain-Computer Interface (BCI) (pp. 1-4). IEEE.

28. C.-C. Fan, H. Yang, Z.-G. Hou, Z.-L. Ni, S. Chen, and Z. Fang, "Bilinear neural network with 3-D attention for brain decoding of motor imagery movements from the human EEG," Cognitive Neurodynamics, vol. 15, pp. 181-189, 2021.

29. W. Tao, Z. Wang, C. M. Wong, Z. Jia, C. Li, X. Chen, C. P. Chen, and F. Wan, "ADFCNN: Attention-Based Dual-Scale Fusion Convolutional Neural Network for Motor Imagery Brain-Computer Interface," IEEE Transactions on Neural Systems and Rehabilitation Engineering, vol. 32, pp. 154-165, 2024.

30. H. Li, X. Zhang, Y. Wan, and X. Zhang, "DSA-MFNet: Deep-shallow Attention based Multi-frame Fusion Network for EEG motor imagery classification." pp. 2021-2026.

31. A. Vaswani, N. Shazeer, N. Parmar, J. Uszkoreit, L. Jones, A. N. Gomez, Ł. Kaiser, and I. Polosukhin, "Attention is all you need." Advances in neural information processing systems, vol. 30,  pp. 5998-6008, 2017

32. A. Dosovitskiy, L. Beyer, A. Kolesnikov, D. Weissenborn, and N. Houlsby, "An Image is Worth 16x16 Words: Transformers for Image Recognition at Scale," In International Conference on Learning Representations, 2020.

33. Z. Liu, Y. Lin, Y. Cao, H. Hu, Y. Wei, Z. Zhang, S. Lin, and B. Guo, "Swin transformer: Hierarchical vision transformer using shifted windows." In Proceedings of the IEEE/CVF international conference on computer vision, pp. 10012-10022, 2021

34. H. Zhang, X. Zhao, Z. Wu, B. Sun, and T. Li, "Motor imagery recognition with automatic EEG channel selection and deep learning," Journal of Neural Engineering, vol. 18, no. 1, pp. 016004, 2021.

35. P. Autthasan, R. Chaisaen, T. Sudhawiyangkul, P. Rangpong, S. Kiatthaveephong, N. Dilokthanakul, G. Bhakdisongkhram, H. Phan, C. Guan, and T. Wilaiprasitporn, "MIN2Net: End-to-end multi-task learning for subject-independent motor imagery EEG classification," IEEE Transactions on Biomedical Engineering, vol. 69, no. 6, pp. 2105-2118, 2021.

36. K. K. Ang, Z. Y. Chin, C. Wang, C. Guan, and H. Zhang, "Filter bank common spatial pattern algorithm on BCI competition IV datasets 2a and 2b," Frontiers in Neuroscience, vol. 6, pp. 39, 2012.

37. S. Ioffe, and C. Szegedy, "Batch normalization: Accelerating deep network training by reducing internal covariate shift." In International conference on machine learning, pp. 448-456, 2015.

38. J. L. Ba, J. R. Kiros, and G. E. Hinton, "Layer normalization," arXiv preprint arXiv:1607.06450, 2016.

39. D. Hendrycks, and K. Gimpel, "Gaussian error linear units (gelus)," arXiv preprint arXiv:1606.08415, 2016.

40. R. Hadsell, S. Chopra, and Y. LeCun, "Dimensionality reduction by learning an invariant mapping." In 2006 IEEE computer society conference on computer vision and pattern recognition (CVPR'06) (Vol. 2, pp. 1735-1742). IEEE.

41. C.-Y. Wu, R. Manmatha, A. J. Smola, and P. Krahenbuhl, "Sampling matters in deep embedding learning." In Proceedings of the IEEE international conference on computer vision, pp. 2840-2848.

42. M. Tangermann, K.-R. Müller, A. Aertsen, N. Birbaumer, C. Braun, C. Brunner, R. Leeb, C. Mehring, K. J. Miller, and G. R. Müller-Putz, "Review of the BCI competition IV," Frontiers in neuroscience, vol. 6, pp. 55, 2012.

43. D. P. Kingma, and J. Ba, "Adam: A method for stochastic optimization," arXiv preprint arXiv:1412.6980, 2014.